\documentclass{emulateapj}
\submitted{{\it Submitted for publication in ApJ}}
\usepackage{multirow,color,wrapfig,ulem}
\usepackage {graphicx}

\usepackage{graphics}
\usepackage[dvips]{epsfig}

\newcommand{\manuscript}{paper }
\newcommand{\lcdm}{$\Lambda$CDM }
\newcommand{\mpc}{\rm{Mpc}}

\newcommand{\ly}{{\ifmmode{{\rm Ly}\alpha~}\else{Ly$\alpha$~}\fi}}
\newcommand{\hMpc}{{\ifmmode{h^{-1}{\rm Mpc}}\else{$h^{-1}$Mpc }\fi}}   
\newcommand{\hGpc}{{\ifmmode{h^{-1}{\rm Gpc}}\else{$h^{-1}$Gpc }\fi}}   
\newcommand{\hmpc}{{\ifmmode{h^{-1}{\rm Mpc}}\else{$h^{-1}$Mpc }\fi}}  
\newcommand{\hkpc}{{\ifmmode{h^{-1}{\rm kpc}}\else{$h^{-1}$kpc }\fi}}  
\newcommand{\hMsun}{{\ifmmode{h^{-1}{\rm
        {M_{\odot}}}}\else{$h^{-1}{\rm{M_{\odot}}}$}\fi}}   
\newcommand{\hmsun}{{\ifmmode{h^{-1}{\rm
        {M_{\odot}}}}\else{$h^{-1}{\rm{M_{\odot}}}$}\fi}}   
\newcommand{\Msun}{{\ifmmode{{\rm {M_{\odot}}}}\else{${\rm{M_{\odot}}}$}\fi}}  
\newcommand{\msun}{{\ifmmode{{\rm {M_{\odot}}}}\else{${\rm{M_{\odot}}}$}\fi}}

\newcommand{\rand}{{\ifmmode{{\mathcal{R}}}\else{${\mathcal{R}}$ }\fi}}  
  
\newcommand{\kms}{{\ifmmode{{\mathrm{\,km\ s}^{-1}}}\else{\,km~s$^{-1}$}\fi}}

\def\lsim{~\rlap{$<$}{\lower 1.0ex\hbox{$\sim$}}}

\def\gsim{~\rlap{$>$}{\lower 1.0ex\hbox{$\sim$}}}

\shorttitle{The Local Group in the Cosmic Web}
\shortauthors{Forero-Romero \& Gonz\'alez}

\begin{document}

\title{The Local Group in the Cosmic Web}
\author{J. E. Forero-Romero$^1$ and R. Gonz\'alez$^{2,3}$}
\affil{$^1$ Departamento de F\'{i}sica, Universidad de los Andes,
  Cra. 1 No. 18A-10, Edificio Ip, Bogot\'a, Colombia\\
  $^2$ Instituto de Astrof\'{i}sica, Pontificia Universidad Cat\'olica de Chile,
  Av. Vicu\~na Mackenna 4860, Santiago, Chile\\  
  $^3$ Centro de Astro-Ingenier\'{i}a, Pontificia Universidad Cat\'olica de Chile,
  Av. Vicu\~na Mackenna 4860, Santiago, Chile
}
\email{je.forero@uniandes.edu.co}
\email{regonzar@astro.puc.cl}

\begin{abstract}
We explore the characteristics of the
cosmic web around Local Group(LG) like pairs  using a cosmological simulation in the
$\Lambda$CDM cosmology.  
We use the Hessian of the gravitational potential to classify regions
on scales of $\sim 2$ Mpc as a peak, sheet, filament or void.  
The sample of LG counterparts is represented by two samples of halo
pairs. 
The first is a general sample composed by pairs with
similar masses and isolation criteria as observed for the LG.  
The second is a subset with additional observed kinematic constraints 
such as relative pair velocity and separation. 
We find that the pairs in the LG sample with all constraints are: (i)
Preferentially   located in filaments and sheets, (ii) Located in in a
narrow range of local overdensity $0<\delta<2$, web ellipticity $0.1<e<1.0$ and
prolateness $-0.4<p<0.4$.  (iii) Strongly aligned with the cosmic web.
The alignments are such that the pair orbital angular momentum tends to
be perpendicular to the smallest tidal eigenvector, $\hat{e}_3$, which
lies along the filament direction or the sheet plane.
A stronger alignment is present for the vector linking the two halos with the vector $\hat{e}_3$. Additionally, we fail to find a strong correlation of the spin of each halo in the pair with the cosmic web.
All these trends are expected to a great extent from the selection on
the LG total mass on the general sample. 
Applied to the observed LG, there is a potential conflict between
the alignments of the different planes of satellites and the numerical
evidence for satellite accretion along filaments; the direction
defined by $\hat{e}_3$. This highlights the relevance of achieving a
precise characterization of the place of the LG in the cosmic web in
the cosmological context provided by $\Lambda$CDM.

\end{abstract}


\section{Introduction}
\label{sec:intro}

The spatial and kinematic configuration of the Local Group
(LG) galaxies is very uncommon in the local Universe and in cosmological
simulations. 
The LG is dominated by two big spirals: the Milky Way (MW) and M31, the next
most-luminous galaxy is M33 which is $\sim 10$ times less massive than
M31, followed by several dozen less luminous dwarf galaxies, up to a
distance of $\sim 3$\mpc.   
The velocity vector of M31, with a low
tangential velocity is consistent with a head-on collision orbit toward
the MW
\citep{2008MNRAS.386..461C,2012ApJ...753....8V,2012ApJ...753....7S}.   

Another feature of the Local Group is the relatively low velocity
dispersion of nearby galaxies up to $\sim 8$ Mpc \citep[][and
  references therein]{1975ApJ...196..313S,2011MNRAS.415L..16A}. 
The environment around the Local Group has density quite close to the
average density of the universe
\citep{2003ApJ...596...19K,2005AJ....129..178K}. 
In addition, the closest massive galaxy cluster, the Virgo Cluster, is
$\sim 16.5\ \mpc$ away \citep{2007ApJ...655..144M}.   

In addition, the LG is located in a diffuse and
warped filament/wall connecting Virgo Cluster with Fornax Cluster, some
nearby galaxies and groups members of this large structure are the
Maffei group, NGC $6744$, NGC $5128$, M$101$, M$81$, NGC$1023$, Cen A
group \citep{2013AJ....146...69C}.
At this scale, there is no evident alignment of the MW-M31 orbital
plane with any local filament or the Virgo-Fornax direction. However,
if we look in a smaller volume below scales of $\sim 6$ \mpc, there is
a clear alignment of the MW-M31 orbit with a local plane of galaxies
as shown by Figure $3$ in \citet{2013AJ....146...69C}. On top of that,
the satellite galaxies in the MW and M31 present different kinds of
strong alignments along planes \citep{Pawlowski2013,Shaya2013}, which
are sometimes considered as unusual in the context of the $\Lambda$
Cold Dark Matter (CDM) model \citep{Pawlowski2012}.

This combination of features makes LG analogues uncommon. 
Using numerical simulations \citet{lganalogues} found less than
$2\%$ MW-sized halos reside in a pair similar to MW-M31 and in a
similar environment. 
Furthermore, if we select pairs constrained
within $2\sigma$ error from current observational measurements of the
velocity components and distance to M31, there are only $46$ systems
in a cubic volume of $250$ \hmpc side, giving a number density $\sim
1.0\times 10^{-6}$Mpc$^{3}$, comparable to the abundance of massive
clusters. 
A similar abundance was found by \cite{ForeroRomero2011} by comparing the
formation history of LG pairs in constrained simulations with the
results of unconstrained cosmological simulations.

\citet{2013ApJ...767L...5F} also studied MW-M31 pairs in numerical
simulations finding the typical quantities characterizing the orbital
parameters of the LG are rare among typical pairs, but not sufficiently rare to
challenge the \lcdm model.  
Another set of criteria for LG analogues was used
by \citet{2008MNRAS.384.1459L}, but despite differences in the
definitions and resulting fraction of LG analogues, results are in
agreement with a low pairs frequency as well.

To better understand the properties of the LG and how this uncommon
pair configuration can be explained in the cosmological context, 
some questions arise. What else can we say of the environment of the
LG on larger scales? To what extent is this an expected configuration
in $\Lambda$CDM? In particular, which are the typical/preferred locations
of these systems within the Cosmic Web? Are the preferential
alignments of satellites a result of the location of the LG in the
Cosmic Web?

In this \manuscript we address those questions by studying the large
scale environment of LG  analogues in the context of \lcdm. 
We use the Bolshoi simulation to explore in what structures they
reside and if there is any correlation or alignment with the cosmic
web.  The large scale environment is defined by the cosmic web
components identified by \citet{Tweb}, and we use the LG analogues computed by
\citet{lganalogues}.

This \manuscript is organized as follows. 
In Section \ref{sec:simulation}
we present the N-body cosmological simulation and the algorithm to
define the cosmic web, next in Section \ref{sec:lg_analogues} we describe the
sample of LG analogues extracted from the simulation. In
Section \ref{sec:results} we present our results to wrap up with a
discussion and conclusions in Section
\ref{sec:discussion}.

\section{Simulation and web finding algorithm}
\label{sec:simulation}

\subsection{The Bolshoi simulation}
We use the Bolshoi simulation of $\Lambda$CDM cosmology: $\Omega_{\rm
  m}=1-\Omega_{\Lambda}=0.27$, $H_0=70\,\rm km/s/Mpc$,
$\sigma_8=0.82$, $n_s=0.95$ \citep{2011ApJ...740..102K}, compatible
with the constraints from the WMAP satellite
\citep{hinshaw_etal13}. The simulation followed the evolution of dark
matter in a $250 \hmpc$ box with spatial resolution of $\sim
1h^{-1}$ kpc and mass resolution of $m_{\rm p}=1.35\times 10^8\ \rm
M_{\odot}$. Halos are identified with the BDM algorithm
\citep{1997astro.ph.12217K}. The BDM algorithm is  a spherical
overdensity halo finding algorithm and is designed to identify both
host halos and subhalos.

\subsection{Cosmic web identification}
The web finding algorithm is based on the tidal tensor computed as the
Hessian of the  gravitational potential field

\begin{equation}
T_{ij} = \frac{\partial^2 \phi}{\partial r_i \partial r_j}, 
\end{equation}
where $r_{i}$, $i=1,2,3$ refers to the three spatial comoving
coordinates and $\phi$ is the gravitational potential renormalized to
follow the Poisson equation $\nabla^2\phi=\delta$ where
$\delta$ is the matter overdensity.  

This tensor is real and symmetric, which means that it can be
diagonalized. 
We denote its eigenvalues as $\lambda_1\geq \lambda_2\geq
\lambda_3$ and their corresponding eigenvectors $\hat{e}_1$,
$\hat{e}_2$ and $\hat{e}_3$. 
The web classification compares each one
of the three eigenvalues to a threshold value $\lambda_{\rm th}$. 
If
the three, two, one or zero eigenvalues are larger than this threshold
the region is classified as peak, filament, sheet or void,
respectively.  Because this tensor is also known as the tidal tensor
we refer to it as the Tweb algorithm.

\cite{Tweb} performed a detailed study for the topology of the
cosmic web and its visual counterpart as a function of the parameter
$\lambda_{\rm th}$. 
They found that reasonable results in terms of the
volume fraction occupied by voids, the visual inspection and the halo
populations in each web type can be reached by values of $0.2<\lambda_{\rm
th}<0.4$. 
In this \manuscript we choose the value of $\lambda_{\rm
  th}=0.3$ to proceed with our analysis. 
This is only relevant to the classification of the simulation into web
elements. Other results are completely independent of this
choice. Nevertheless we have checked that the main conclusions of this
work do not depend on the precise choice of $\lambda_{\rm th}$.

The algorithm to compute the potential is grid based. 
First we interpolate the mass into a cubic grid with a
Cloud-In-Cell (CIC) scheme and smooth it with a Gaussian kernel in order to reduce the grid influence in the computations that follow. 
Then we obtain the gravitational potential using FFT methods and use finite differences
to compute the Hessian at every point in the grid. 
In our case we have used a grid size and a Gaussian smoothing with
two times larger as the typical separation between the two halos in the Local Group. 
The purpose of this choice is to have both halos in the pair a common
environment. 
In this \manuscript we use a grid spacing of $s=0.97$ \hMpc,
corresponding to a $256^3$ grid in the Bolshoi volume. 
The scale for the Gaussian smoothing uses the same value. 

We use the matter overdensity, ellipticity and the prolatenes to
further characterize the web. 
These quantities are defined in terms of the
eigenvalues as follows 
\begin{equation}
\delta = \lambda_1 + \lambda_2 + \lambda_3,
\end{equation}
\begin{equation}
e= \frac{\lambda_3 - \lambda_1}{2(\lambda_1 + \lambda_2 + \lambda_3)}, 
\end{equation}
\begin{equation}
p= \frac{\lambda_1 + \lambda_3 - 2\lambda_2}{2(\lambda_1 + \lambda_2 +
  \lambda_3)}.
\end{equation}

We also measure the alignment of the LG halos with respect to the
cosmic web defined by their eigenvectors.
To this end we characterize each LG pair by two vectors. 
The first is $\hat{n}$, the axis along the orbital angular
momentum of the pair, normal to its orbital plane; the second is
$\hat{r}$, the vector that connects the halos in the pair which can be
related to the alignment of the radial velocities to the web.
We quantify the alignment using the absolute value of the cosine of
the angle between the two vectors of interest $\mu=|\hat{e}_i \cdot
\hat{n}|$ or $\mu=|\hat{e}_i\cdot \hat{r}|$, where $i=1,2,3$.

We have
veryfied that the main trends reported in this paper remain unchanged
if we use the results from a $512^3$ grid smoothed over scales of
$s=0.48$\hMpc. Small changes of a factor of $\sim2$ in the smoothing
scale do not significantly impact the cosmic web as it 
was shown in \cite{Tweb} from the study of the volume and mass
filling fractions for different smoothing scales. Other works have
also shown that broad features the cosmic web, as defined in this
paper, are robust to factor of $\sim2$ changes on the scale defining the web
\citep{Cautun2014}.    

The data of the BDM halos and the Tweb is publicly available through
a database located at \url{http://www.cosmosim.org/}. A detailed
description of the database structure was presented by \cite{Riebe2013}.

\section{Local Group Analogues}
\label{sec:lg_analogues}

To construct a sample of the MW-M31 pairs at $z\sim 0$, we use a
series of simulation snapshots  at $z<0.1$ (since the last $\sim
1.3$ Gyr) spaced by $\sim 150-250$ Myr. 
This is done because a particular configuration of MW and M31 is transient and
corresponds to a relatively small number of systems at one
snapshot. 
By using multiple snapshots we can increase the sample of systems in
such configuration during a period of time in which secular
cosmological evolution is small.  

The LG analogues or the General Sample (GS) in this paper are pairs selected in
relative isolation, and in a wide range of masses from  $M_{200c}=5
\times 10^{11}$ \msun $ $ to $ 5 \times 10^{13}$ \msun.  
Isolation criteria include pair members closer than $1.3$\mpc,
 and with no massive neighbors within $5$\mpc.
In addition, we require that pairs have no Virgo-like neighbor halo
with mass $M_{200c}>1.5 \times 10^{14}\ \rm M_{\odot}$ within $12$
Mpc.  
We have $5480$ pairs under these general criteria. 
A full description of the selection criteria can be found in
\citet{lganalogues,sat}.   

We also define two subsets of restricted samples (RS) more closely
related to the MW-M31  dynamics according to the tolerance in
additional constraints. 
A sample named $2\sigma$, corresponding to LG analogues constrained by
two times the observational errors in the orbital values (radial
velocity, tangential velocity, and separation), and a more relaxed
sample named $3\sigma$ for LG analogues constrained by three times
observational errors accordingly.  
The number of pairs in each sample is $46$ and $120$ respectively,
notice we have less pairs than in \citet{lganalogues}, since
we removed pairs which are too close at $z=0$, i.e. their virial radii
overlap, also we removed a couple of pairs that merged or change their
mass more than $20\%$ at present time since they were detected at
$z<0.1$. 

\section{Results}
\label{sec:results}

\begin{figure}
\begin{center}
  \includegraphics[width=0.50\textwidth]{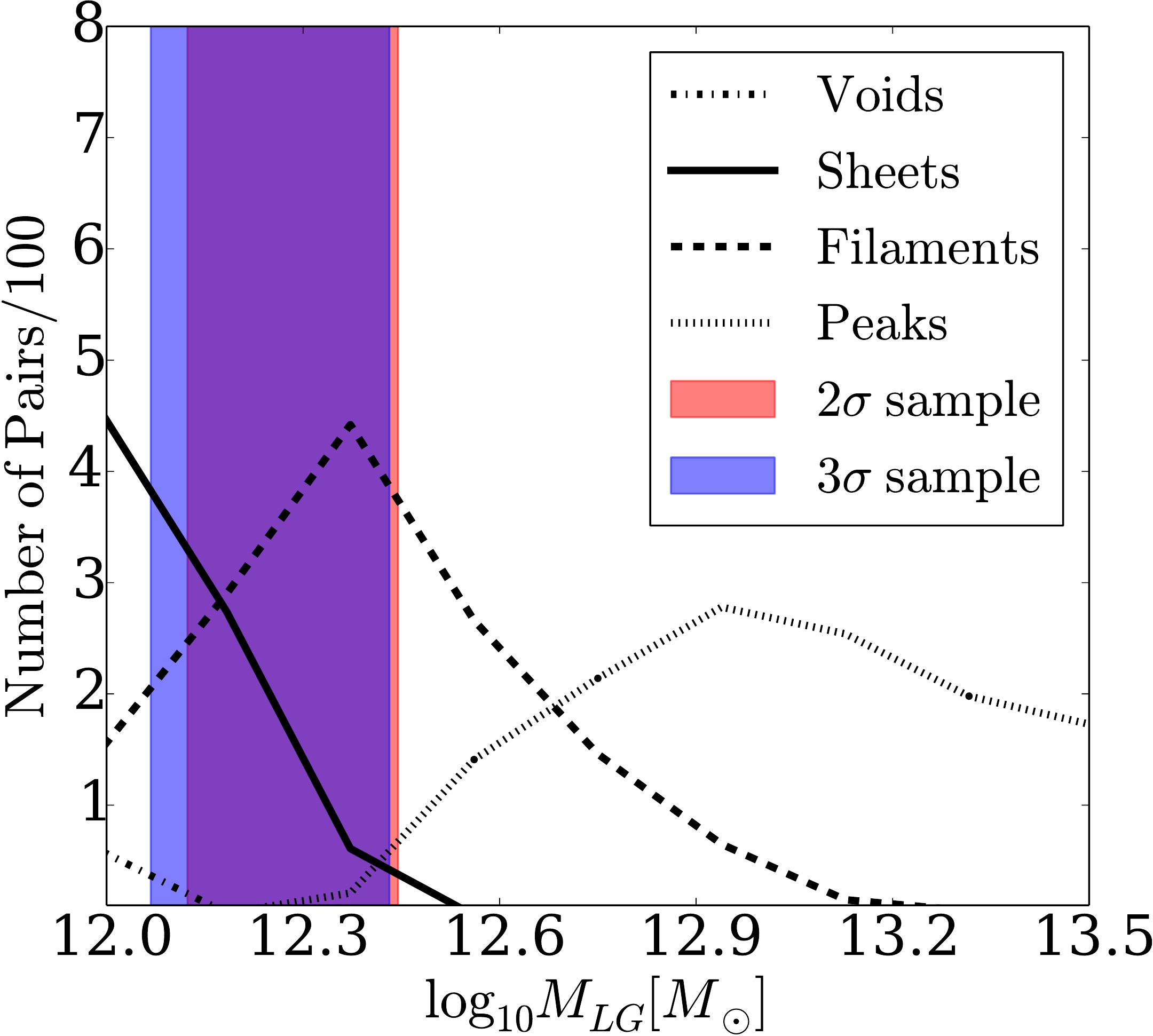}
\caption{Mass distribution of pairs in different environments
for the general sample.
The shaded regions show the $68\%$ confidence interval of the mass ranges for $2\sigma$ and $3\sigma$
samples.  
\label{fig:median_fraction}}
\end{center}
\end{figure}

\begin{table*}
\begin{center}
\begin{tabular}{ccccc}\hline\hline
Sample & Peak & Filament & Sheet & Void\\
       & $n$ (\%) & $n$ (\%) & $n$ (\%) & $n$ (\%) \\\hline
2$\sigma$ & 4 (8.7) & 24 (52.2) &  17 (36.7) & 1 (2.2)\\
3$\sigma$ & 10 (8.3) & 58 (48.3) & 47 (39.2) & 5 (4.2)\\  
General & 1312 (23.9) & 1472 (26.9) & 1769 (32.3) & 927 (16.9)\\
General ($12.1<\log_{10} M_{\rm LG}/\Msun<12.3$)& 8 (1.4) & 334 (55.5) & 259
(43.0) & 1 (0.1)\\
\hline\hline
\end{tabular}
\caption{
Number of pairs in the four different kinds of environments for each
of the three samples presented in Section \ref{sec:lg_analogues}. 
In parenthesis the same number as a percentage of the total population.
The last line in the table corresponds to the general sample with an
additional mass cut for the total pair mass.  
\label{table:web_type}}
\end{center}
\end{table*}

\begin{figure*}
\begin{center}
  \includegraphics[width=0.31\textwidth]{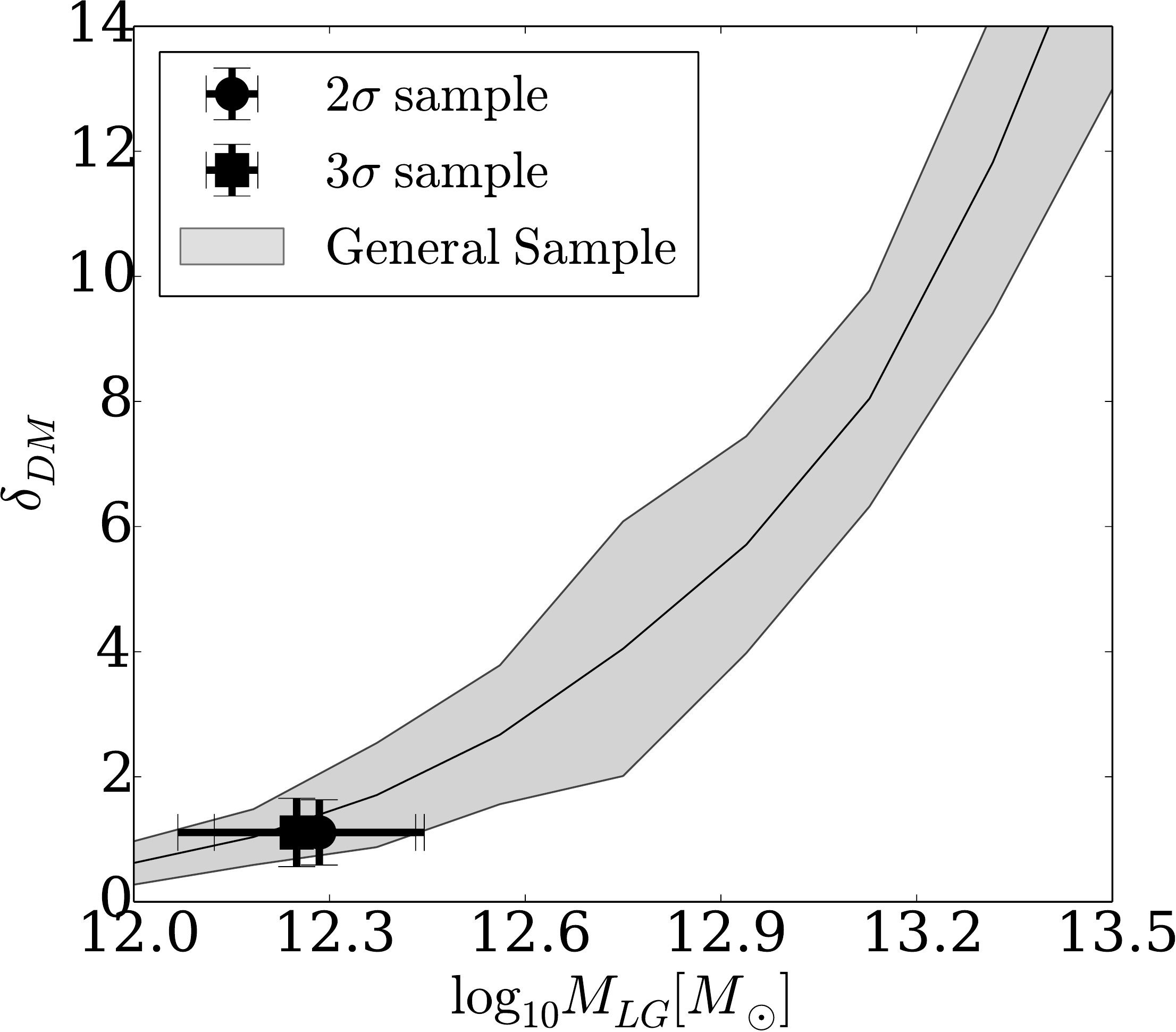}
  \includegraphics[width=0.32\textwidth]{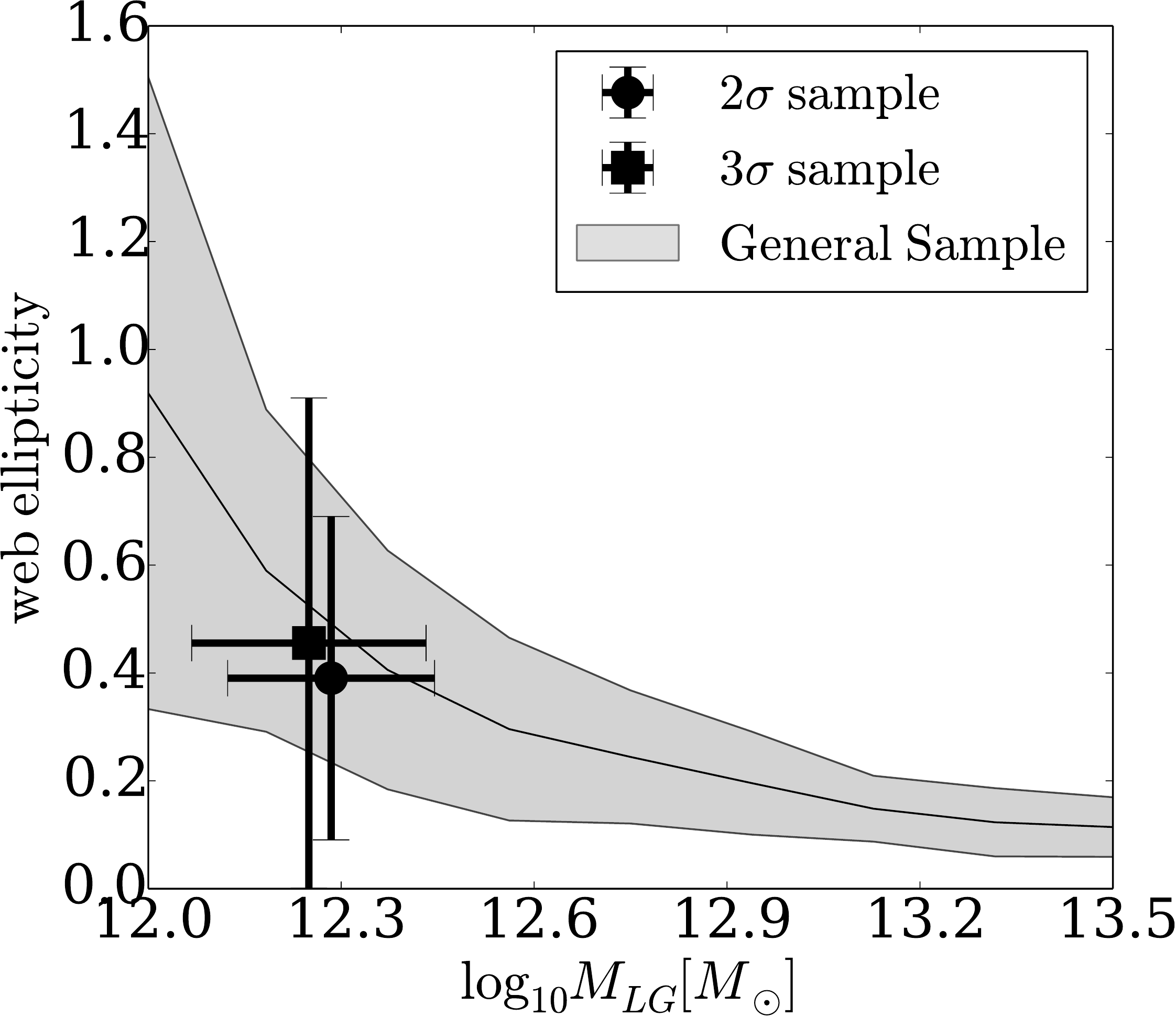}
  \includegraphics[width=0.32\textwidth]{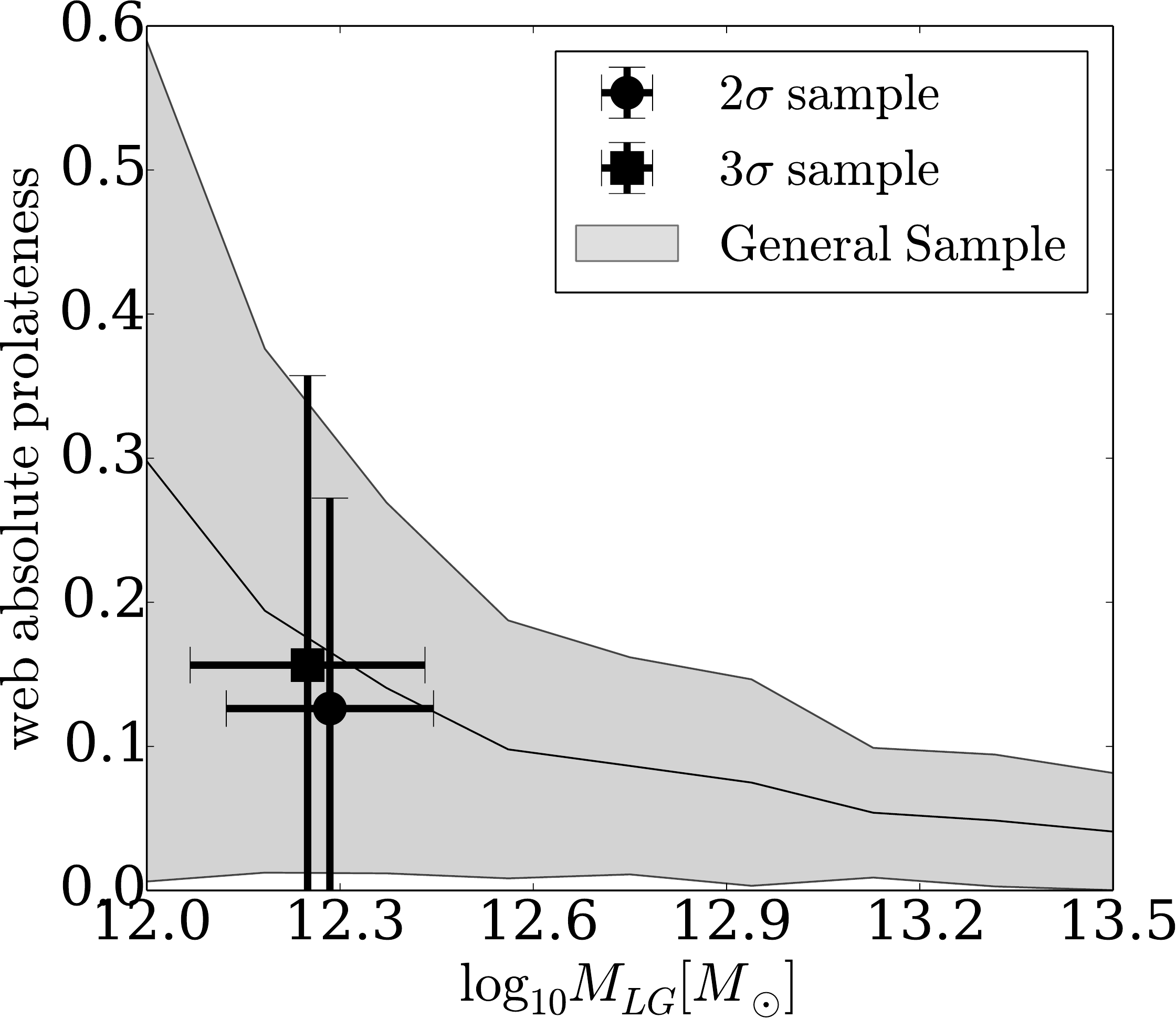}
\caption{Mass dependency of the average dark matter overdensity (left),
  web ellipticity (middle) and web absolute value prolateness (right) at the
  pair location. The shaded contours indicate the $68\%$ confidence region. 
\label{fig:median_overdensity}}
\end{center}
\end{figure*}

\begin{figure*}
\begin{center}
  \includegraphics[width=0.32\textwidth]{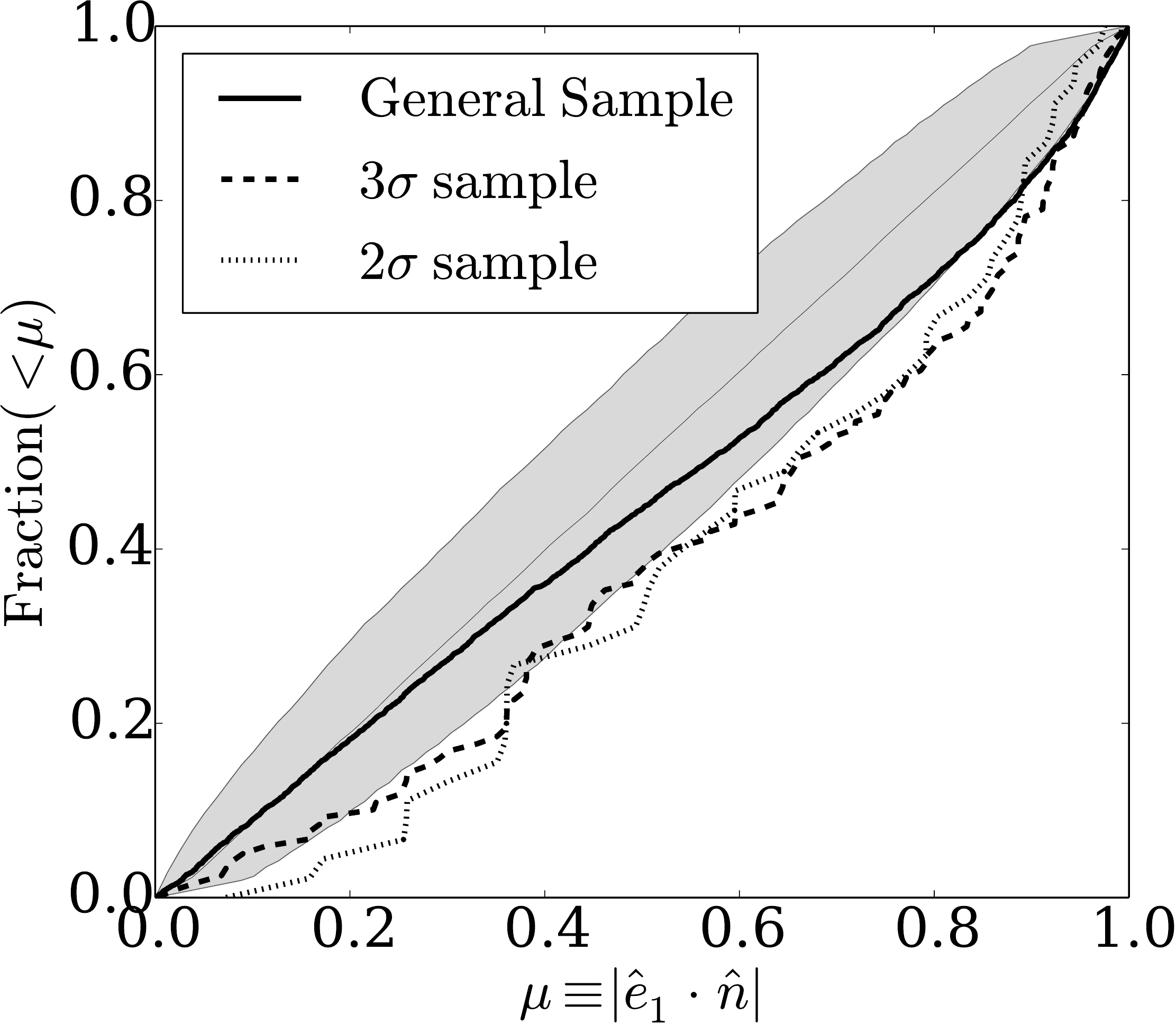}
  \includegraphics[width=0.32\textwidth]{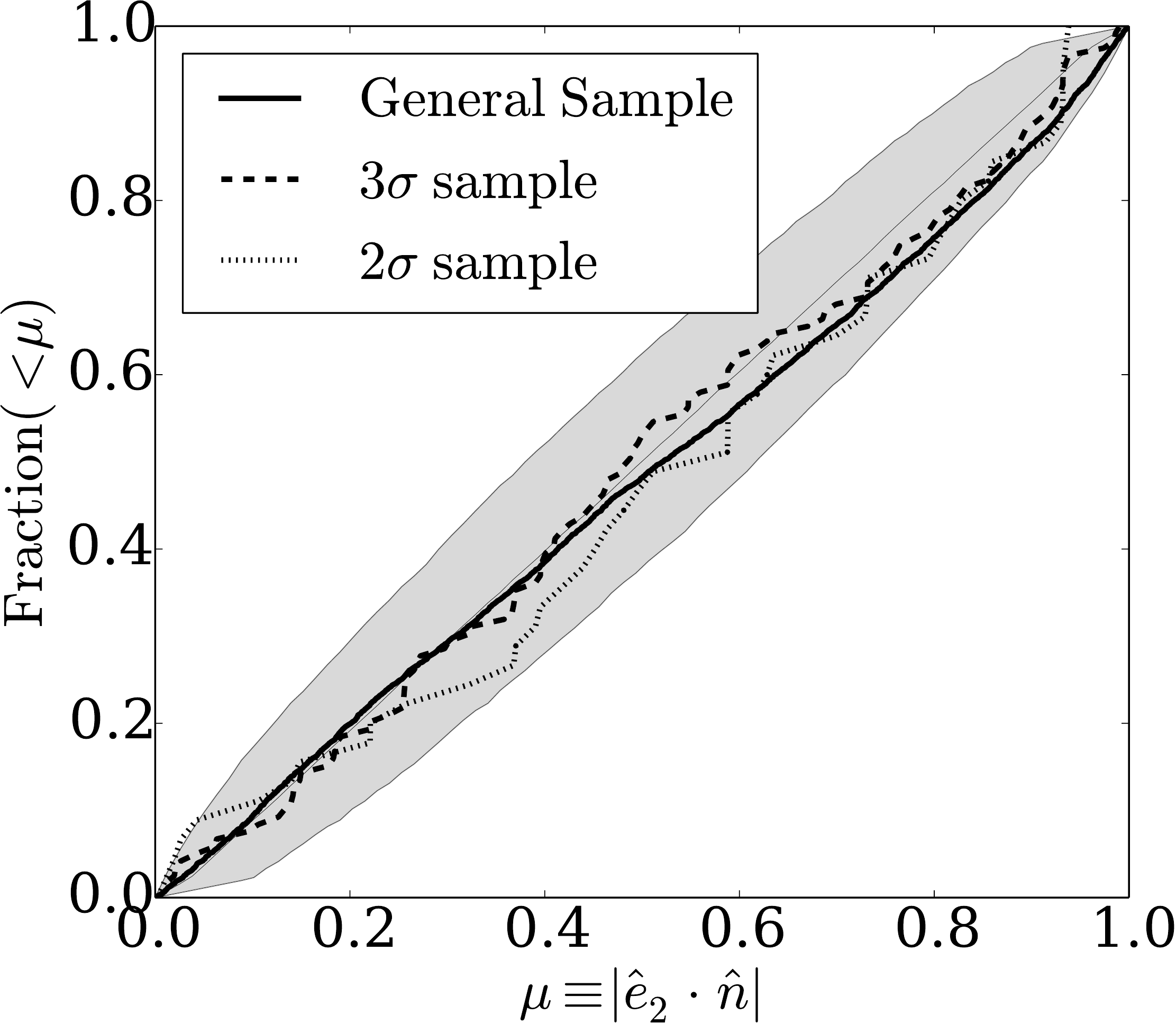}
  \includegraphics[width=0.32\textwidth]{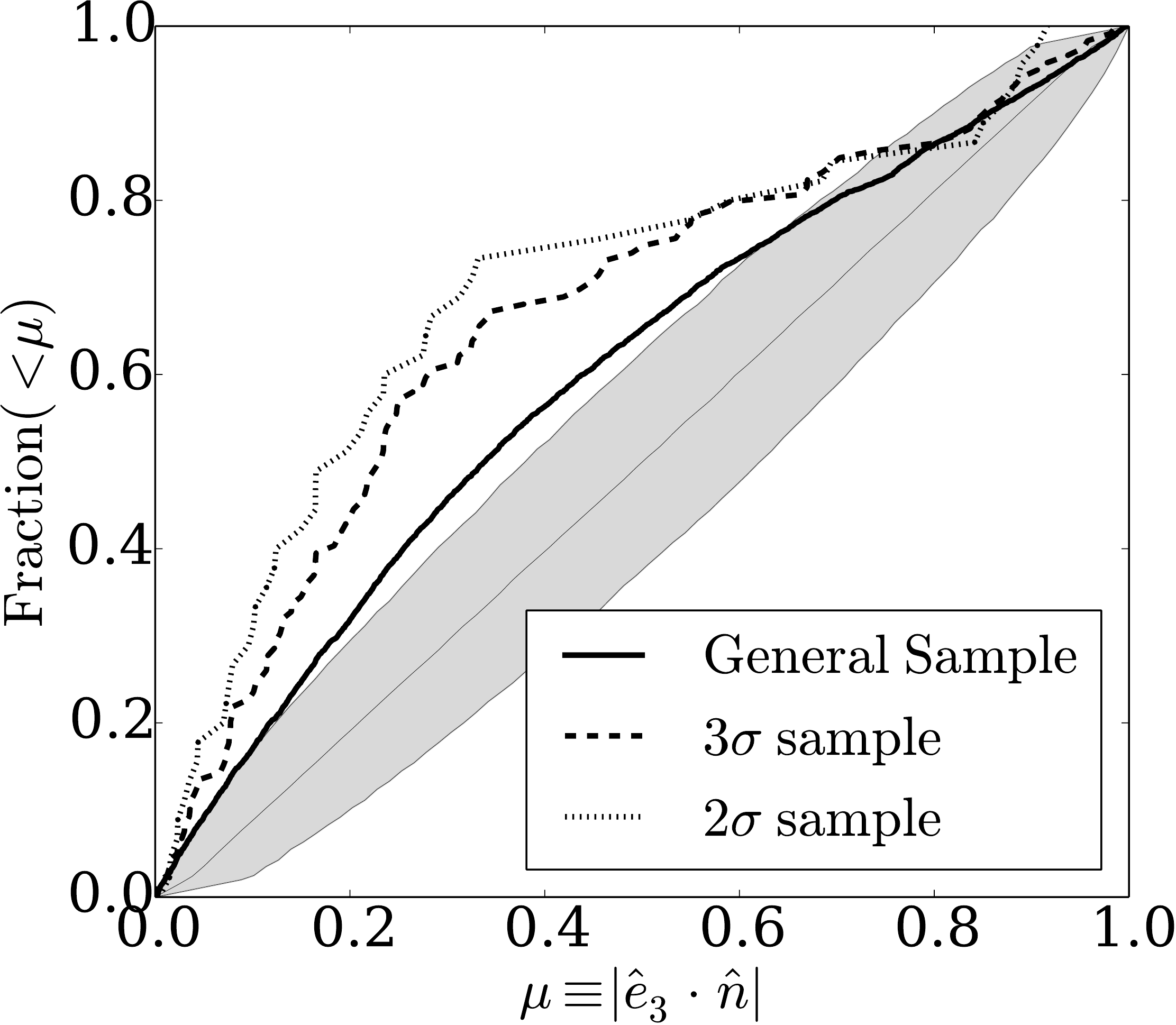}
\end{center}
\caption{Cumulative distributions for the alignment between the normal
  vector to the pair orbital plane, $\hat{n}$, and the three eigenvectors in
  the Tweb. 
The shaded region shows the expectation for a random
  distribution without any preferential alignment, it encloses the $5\%$ and $95\%$ percentiles of 10000
  flat distributions for $\mu$ generated with the  same number of
  points as the $2\sigma$ sample.  
    \label{fig:alignment_n}}  
\end{figure*}

\begin{figure*}
\begin{center}
  \includegraphics[width=0.32\textwidth]{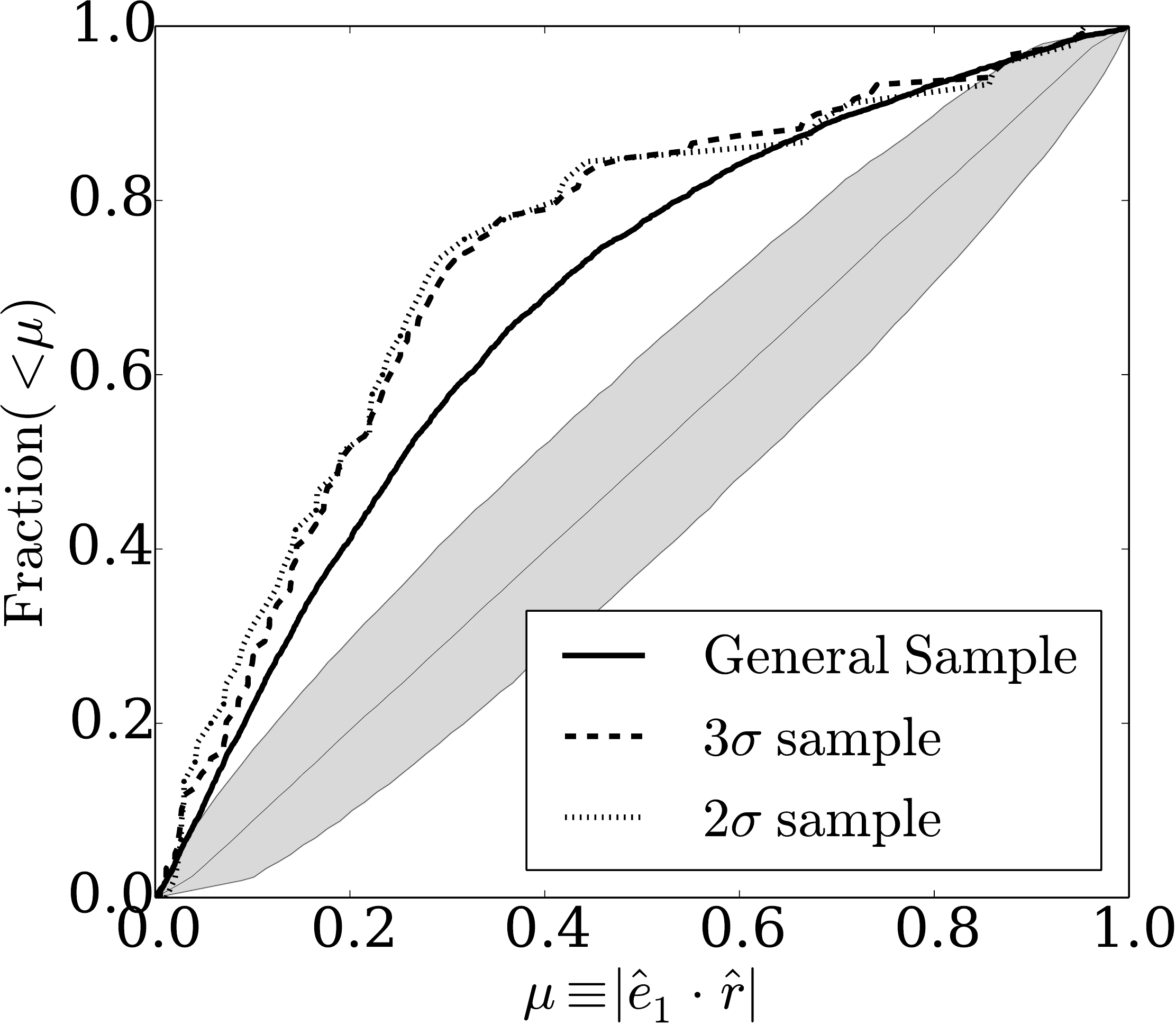}
  \includegraphics[width=0.32\textwidth]{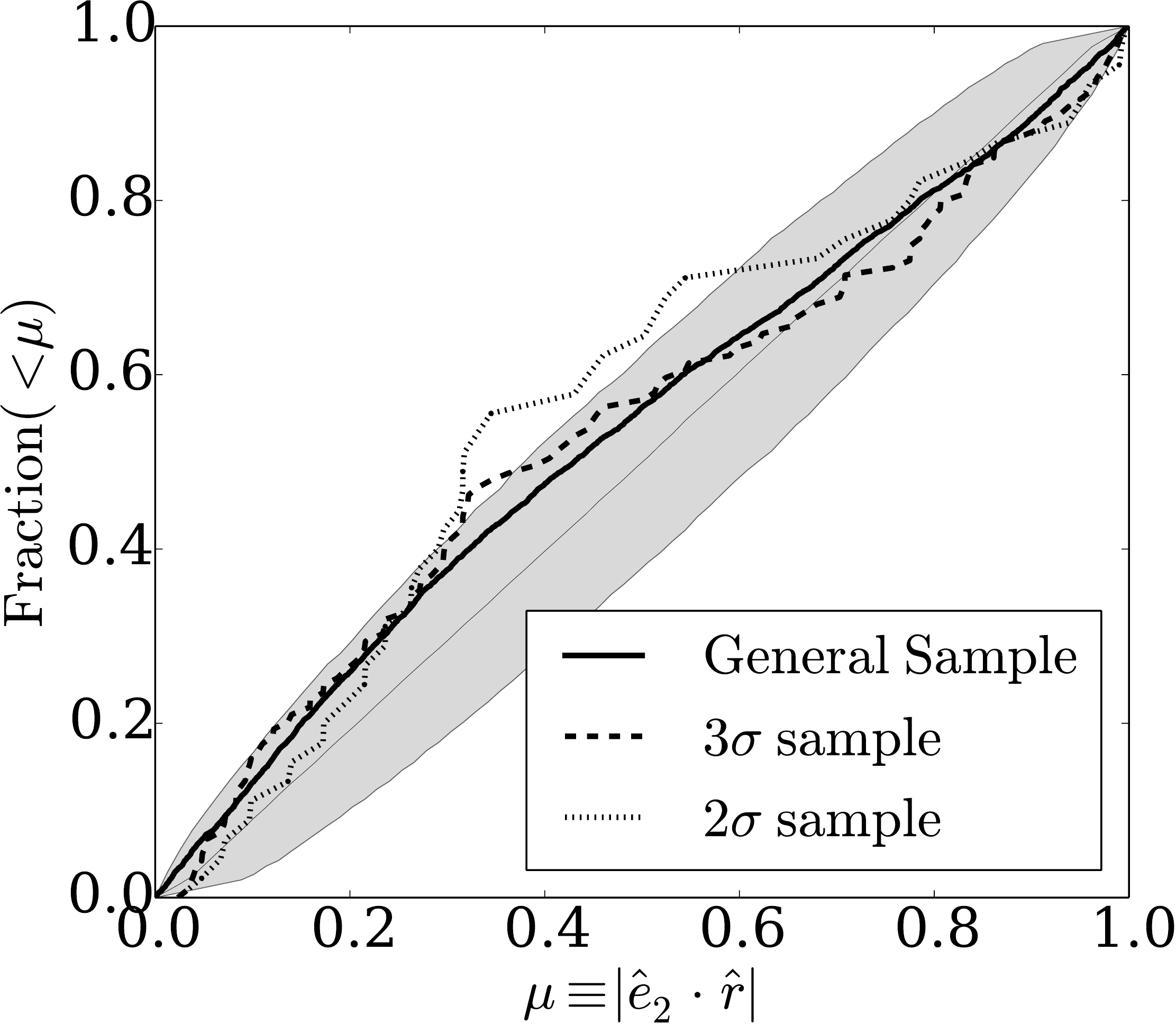}
  \includegraphics[width=0.32\textwidth]{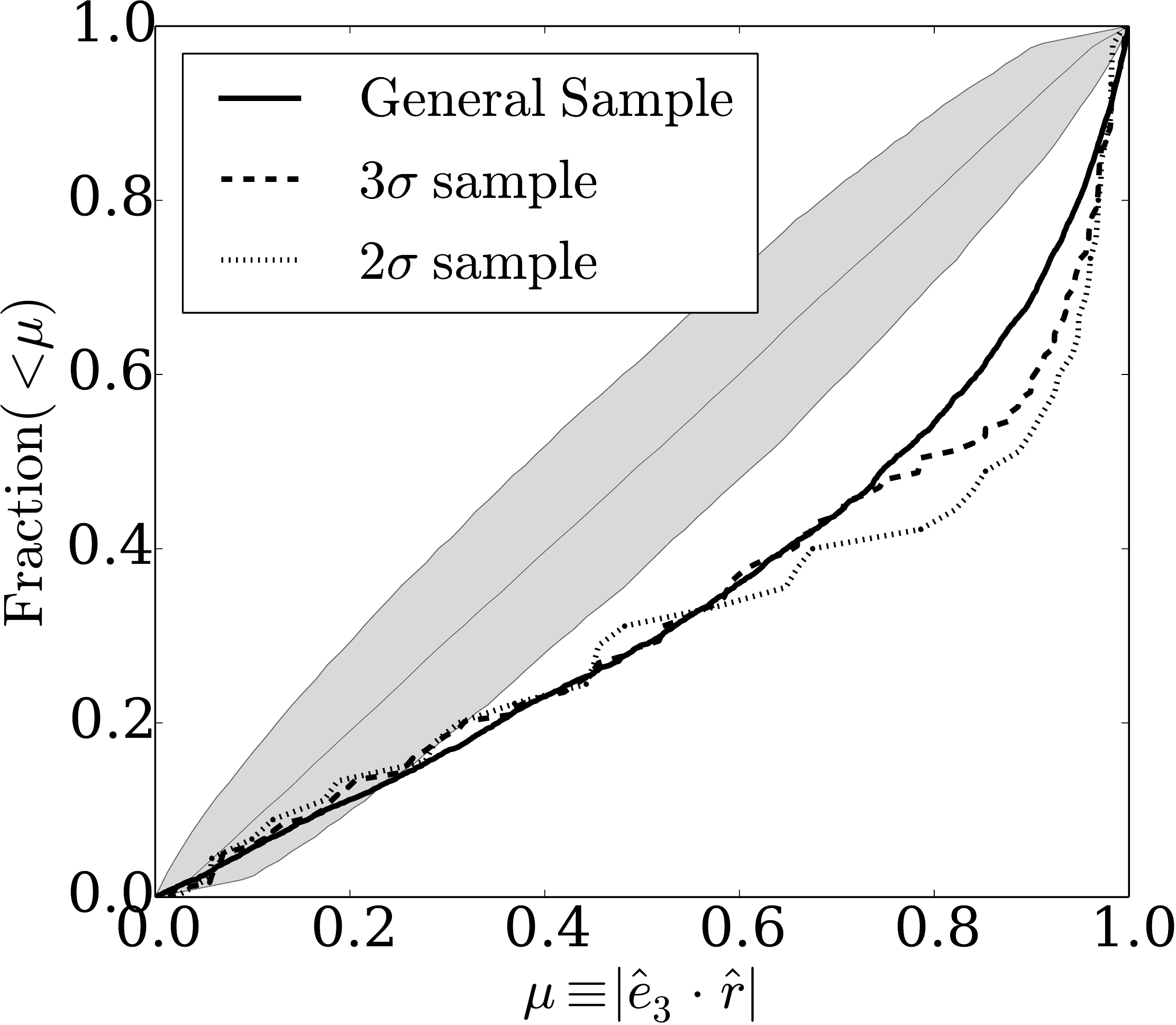}
\end{center}
\caption{Cumulative distributions for the alignment between the vector
  linking the two halos in the pair, $\hat{r}$, and the three
  eigenvectors in the Tweb. The shaded region is the same as in
  Figure \ref{fig:alignment_n}.
    \label{fig:alignment_r}}  
\end{figure*}

\begin{figure*}
\begin{center}
  \includegraphics[width=0.45\textwidth]{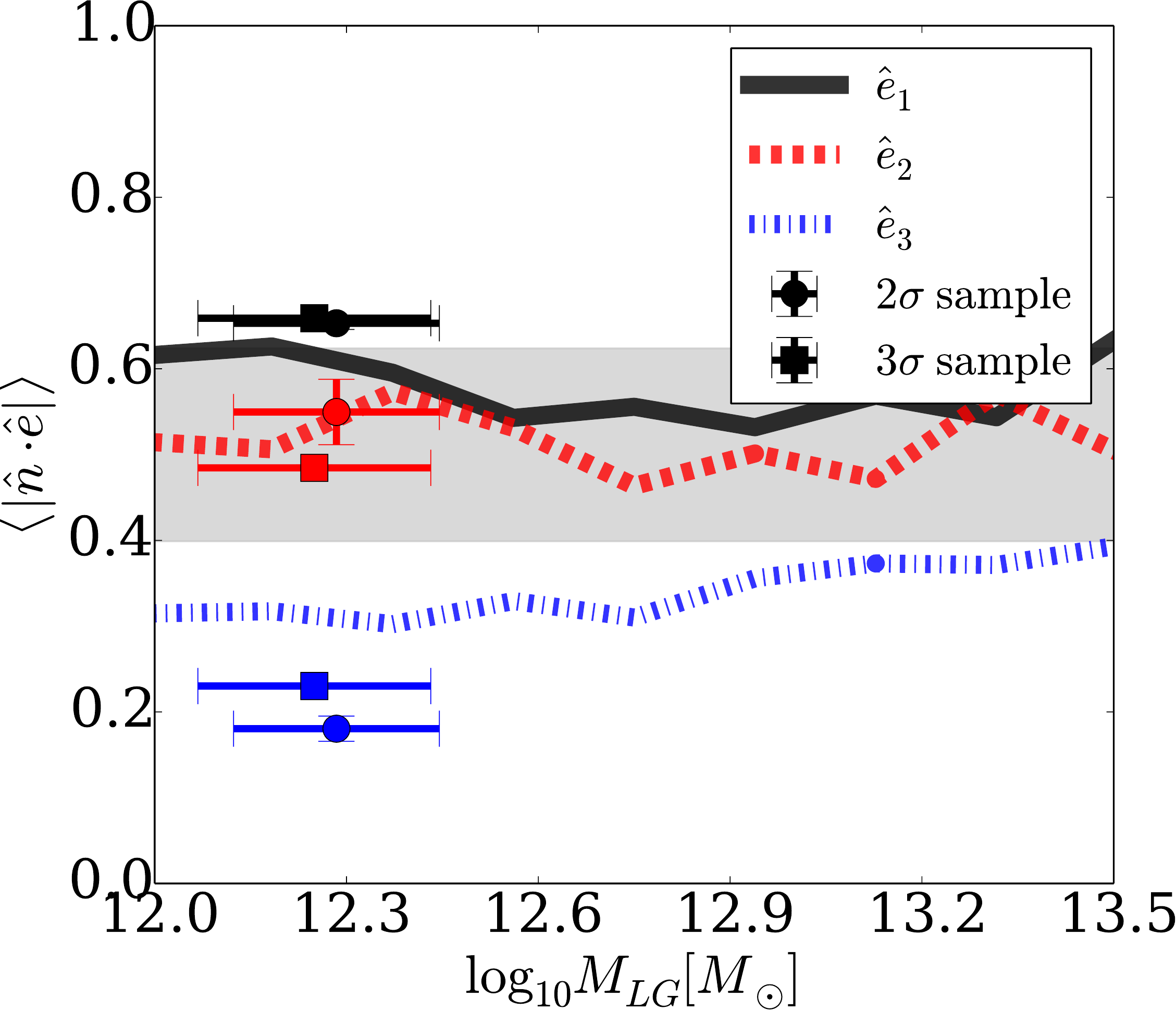}
  \includegraphics[width=0.45\textwidth]{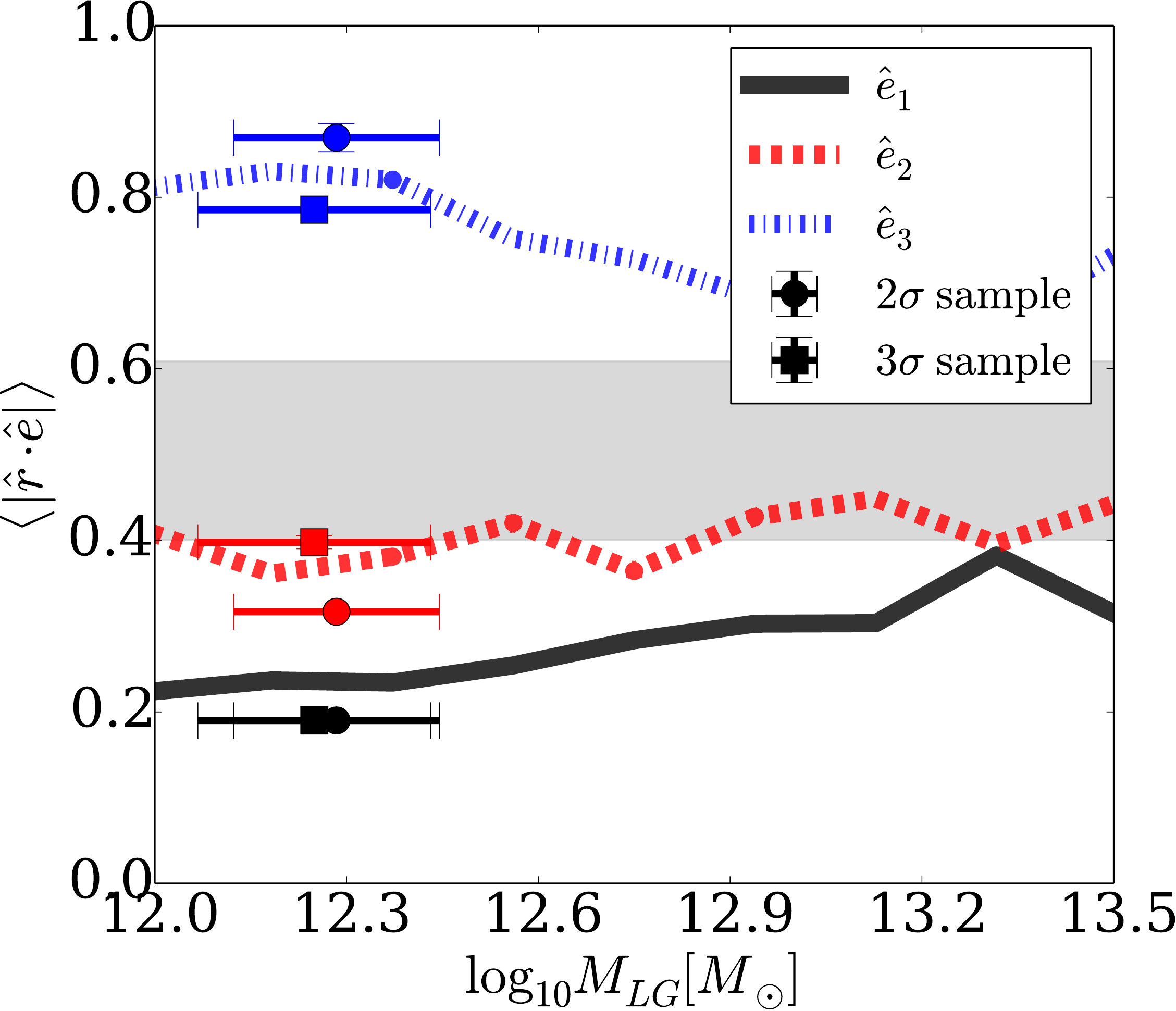}
\caption{Mass dependency of the median value for the dot product
  between the normal/radial vector $\hat{n}$/$\hat{r}$ (left/right)
  and each one of the  eigenvectors.  The lines show the trends for
  the general sample. The error bars in the 2$\sigma$ and 3$\sigma$ points
  correspond to jackknife estimates. 
  The shaded region shows the expectation for a random distribution
 without any preferential alignment, it encloses the $5\%$ and $95\%$
 percentiles of 10000  flat distributions for $\mu$ generated with the
 same number of points as the $2\sigma$ sample.   
\label{fig:median_alignment_n}}
\end{center}
\end{figure*}

\begin{figure}
\begin{center}
  \includegraphics[width=0.45\textwidth]{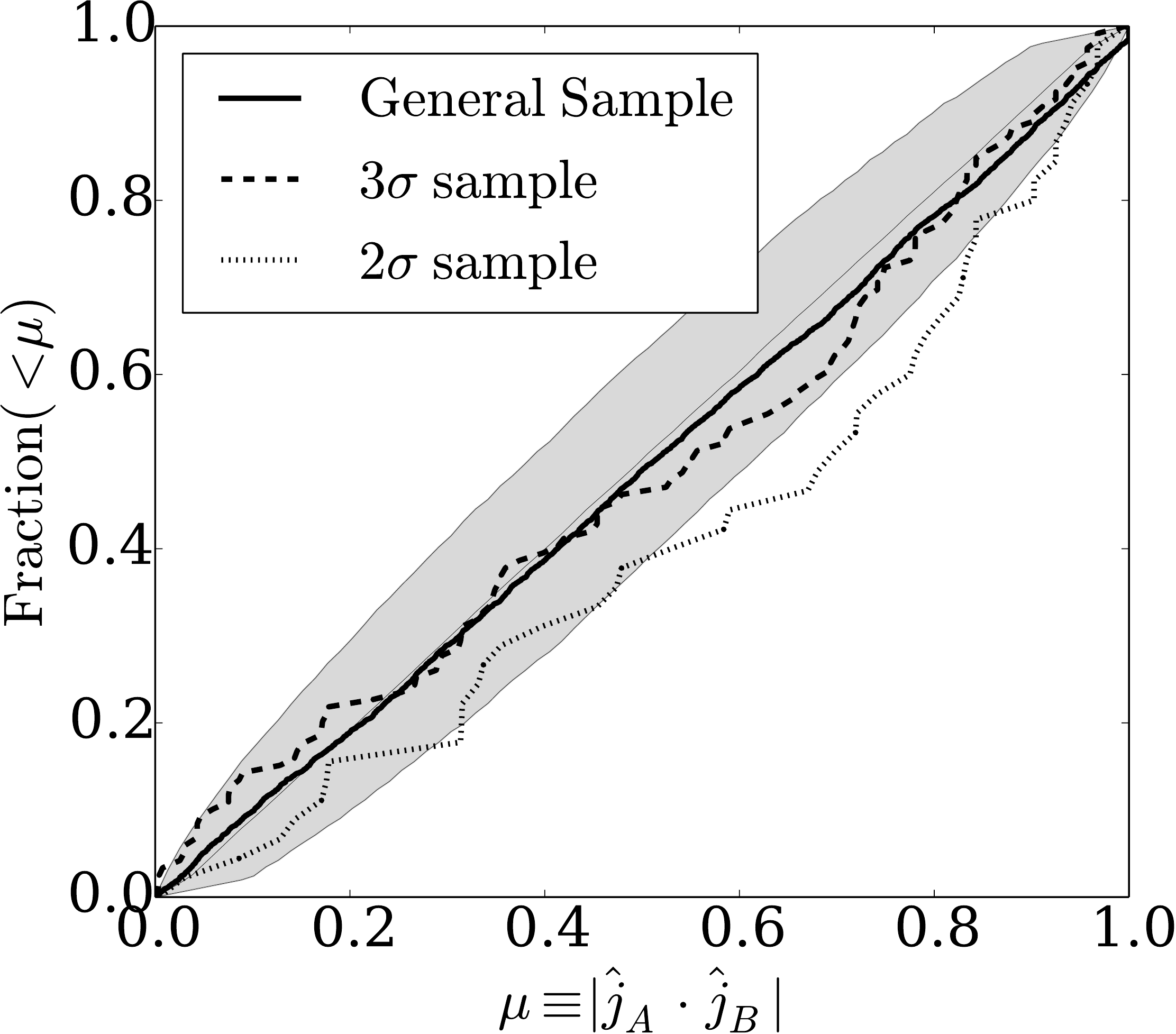}
\end{center}
\caption{Alignment between the two angular momentum vectors of the two
  halos in the pair.
  The shaded region shows the expectation for a random
  distribution without a preferential alignment as it is described in
  Figure \ref{fig:median_alignment_n}.
    \label{fig:jj_alignment}}  
\end{figure}

\subsection{The preferred environment for LGs}

The first result we explore is the kind of environment occupied by
our LGs, and use as reference sample the GS where pairs are much more
relaxed in mass and kinematic constraints.
We find that the pairs in the GS are located across all different
evironment without any strong preferences; $1/3$  are located in
sheets, $1/4$ in peaks, $1/4$ in filaments and the remaining $1/6$ in
voids.  

The situation in the $2\sigma$ and $3\sigma$ samples is very
different. 
By and large the LGs in these samples are located in filaments and sheets. 
In both samples, $\sim 50\%$ of the pairs can be found in filaments
while $\sim 40\%$ are in sheets. 
These absolute numbers in each environment for each sample are
presented in Table \ref{table:web_type}.  

We find that the difference between the GS and the RS 
can be explained by the total pair mass.
In \citet{lganalogues} the mass range covered by $2\sigma$ and $3\sigma$ 
samples is very narrow and it is used to constraint the LG mass.
We show in Table \ref{table:web_type} that a subset of the GS
having a similar mass range than the RS reproduces similar environment
fractions. The correlation between halo mass and their environment is a
well known result \citep{1998ApJ...500...14L}.

Figure \ref{fig:median_fraction} summarizes this correlation between
environment an total pair mass.
Each line represents the mass distribution of pairs in the four
different environments for the GS.
High mass pairs tend to be located in  peaks and filaments while less
massive ones in voids and sheets. 
The shaded regions represent the $68\%$ confidence intervals of the mass 
distributions of $2\sigma$ and $3\sigma$ samples.

\subsection{Web Overdensity, Ellipticity and Prolateness}

We now characterize the preferred place of the pair samples in terms of the
web overdensity, ellipticiy and anisotropy as defined in Section
\ref{sec:simulation}. 

Figure \ref{fig:median_overdensity} shows the dependency of these three
characteristics  on the total pair mass for the different samples.
GS is represented by the solid lines with the associated 
errors covered by the shaded region. 
The symbols represents the results for the $2\sigma$ and $3\sigma$
samples.   
In the three cases it is evident that the range of values for the
$2\sigma$ and $3\sigma$ samples are completely expected from the mass
constraint alone.

The left panel shows the overdensity dependence on pair mass. 
Higher mass pairs are located in high density regions.
The $2\sigma$ and $3\sigma$ samples having a narrower mass range as 
shown in previous figure, are consequently located within a narrower 
range of overdensities $0.0<\delta<2.0$ peaking at $\delta \sim 1$. 
This is also consistent with the fact that these samples are mostly 
found in filaments and sheets. 

Middle and right panels show web ellipticity and absolute prolateness
dependence on mass. 
Again we notice that within the $2\sigma$ and $3\sigma$ mass range,
the average ellipticity and prolateness does not differ significantly
from the GS expectation.
In the RS samples the pairs are located in a narrow range for
ellipticities $0.1<e<1.0$ and absolute prolateness $|p|<0.4$.

\subsection{Alignments with the cosmic web}

We now study different alignments of the LG with respect to the cosmic web. 

{\bf Orbital Angular Momentum}. Figure \ref{fig:alignment_n} shows the
cumulative distribution of $\mu\equiv\hat{e}_i\cdot\hat{n}$  for the
three eigenvectors $i=1,2,3$.    
Lines in each panel correspond to different samples.
The straight line across the diagonal shows the expected
result for vectors with randomly distributed directions.

There are two important features in Figure \ref{fig:alignment_n}.
First the alignments themselves. 
There is a strong anti-alignment signature between $\hat{n}$ and the
third eigenvector. 
With respect to the first and second eigenvector the distribution is consistent
with no alignment. 
Second, the alignment strength changes for the different samples. 
For the anti-alignment with $\hat{e}_3$ the signal strengthens as we
move from the GS to the 3$\sigma$ into the 2$\sigma$ sample.

Quantitatively, the anti-alignment feature found with the $\hat{e}_3$ 
vector means that for 2$\sigma$ sample, $\sim 50\%$ pairs have
$|\mu|<0.2$ ($>78$ degrees angle), and $\sim 75\%$ pairs have $|\mu|<0.4$
($>66$ degrees angle). 
These signals do not change significantly on
different environments as has been already show in different alignment
studies with similar \citep{Libeskind2013} or identical
\citep{ForeroRomero2014} web finding techniques as ours.
In particular, these trends hold for pairs in filaments and walls.
If we consider only pairs in filaments, we have that the pair orbital
angular momentum tends to be perpendicular to the filament direction,
in the case of sheets it tends to lie perpendicular to the sheet plane.

The alignment strength could be explained to a great extend by 
a total mass dependency. In Figure \ref{fig:median_alignment_n} we
show the median of $\mu\equiv\hat{e}_i\cdot\hat{n}$ and $\mu\equiv
\hat{e}_i\cdot\hat{r}$ in different mass bins. 
Lines show the median $\mu$-mass relation for the three
eigenvectors in the GS, dots represent the results for the
2$\sigma$-3$\sigma$ samples.  Around the total mass $1.5-2.0 \times
10^{12}$ \msun in the RS samples, the median of $\mu$ clearly from the
2$\sigma$ and 3$\sigma$ samples clearly differs from the GS
results. However, the uncertainty level on the median for a small
sample (indicated by the shaded region) makes this trend compatible
with the values from the GS sample.

We also explored characteristics in the orbits which could be
responsible of the strengthen alignment feature. We  
have that RS samples constrains the tangential velocity, as a
consequence pair orbits are more eccentric as we tighten pair
constraints. For the GS sample there is no tangential velocity
constraint at all.
We studied the relationship between orbit eccentricity and alignment
computing for each halo pair in all samples an approximate orbit 
eccentricity assuming a two body orbit with masses set as the virial 
masses and initial conditions given by current velocity vector and separation.
We found that  selecting a subsample of GS having the same eccentricity than 
2$\sigma$-3$\sigma$ samples does not increase in the alignment when
compared with the full GS.

{\bf Radial Vector}.  Figure \ref{fig:alignment_r} presents the
results for the eigenvectors alignments with respect to the vector
connecting the two halos. 
In this case we find that the vector $\hat{r}$
is strongly aligned along the direction defined by
$\hat{e}_3$ and anti-aligned along $\hat{e}_1$; correspondingly the
signal along $\hat{e}_2$ is rather weak. 

We also find a stronger signal as we move into more restrictive
samples, although the signal from the GS is already very
significant. 
Quantitatively, the alignment feature with $\hat{e}_3$
means that for the 2$\sigma$ and 3$\sigma$ samples, $\sim 50\%$ pairs
have $|\mu|>0.8$ ($<36$ degrees angle) and $\sim 25\%$ pairs have
$|\mu|>0.95$ ($<18$ degrees angle). Similar to the previous case of
the $\hat{n}$ vector, the increasing strength of the
$\hat{r}$ alignment can be explained to a great extenct by a selection
on mass.  

Considering that the 2$\sigma$ and 3$\sigma$ samples correspond to
pairs moving along the radial direction, we can say that the
motion of the LG halos is mostly done along the $\hat{e}_3$
vectors. This is consistent with recent results that report a strong
alignment of halo's peculiar velocities along that direction
\citep{ForeroRomero2014}. 

{\bf Halo Spin}. We also explore the alignment of the  angular
momentum (spin) of each pair member $\hat{j}_A$ and $\hat{j}_B$ with
each other, the orbital angular momentum and with the cosmic web. 

Figure \ref{fig:jj_alignment} shows the cumulative distribution of dot
product between angular momentum of the  two halos. We find a a slight
alignment of spin vectors for the $2\sigma$ sample with a median around
$|\mu|\sim 0.7$ ($45$ degrees) which is barely above the no-alignment expectation marked by the shaded region. Furthermore, we found no significant
alignment with the pair orbital angular momentum nor the cosmic web. 

The absence of alignment for the spins is consistent with different studies of
spin alingnment that mark the range $10^{12}$\Msun as either a transition mass
from alignment into anti-alignment or from no-alignment into alignment
\citep{Hahn2007,AragonCalvo2007,Zhang2009,Codis2012,Trowland2013,Libeskind2013,ForeroRomero2014}. We refer the reader to Table 2 in \cite{ForeroRomero2014}
for a review on the different results of spin alignment with the
cosmic web in cosmological simulations. It seems that this transition
has to do with the changes on the major merger rate as a dark matter
halo approaches the $10^{12}$\Msun mass \citep{2009MNRAS.399..762F,Codis2012}.

In the LG, the angle between MW and M31 spin is $\sim60^{\circ}$
($|\mu|=0.5$) and the angles between spins and orbital angular
momentum are $\sim33^{\circ}$ ($|\mu|=0.83$) and $\sim76^{\circ}$
($|\mu|=0.24$) for MW and M31 respectively
\citep{2012ApJ...753....9V}. 
This is consistent with our results of no strong alignment.

\section{Discussion and Conclusions}
\label{sec:discussion}

The mass range of the LG pairs is tightly correlated with the
properties of the web in which they reside, as shown in Table
\ref{table:web_type} and Figure \ref{fig:median_fraction}. This
confirms that the local overdensity, the trace of the tidal tensor,
is the dominant web parameter that define the abundance and properties
of halos 
\citep{1998ApJ...500...14L,1999MNRAS.302..111L,2004MNRAS.350.1385S,2009MNRAS.394.1825F,
  Alonso2014}. Other
quantities derived from the tidal tensor play a secondary role
defining properties such as its formation history. 

In our case, the fact that the preferred LG total mass is around
$1-4\times 10^{12}$\Msun $\,$implies that the preferred environment are
filaments and sheets with an overdensity close to the average
value. 
Correspondingly, the values for the ellipticity and prolateness are
also well defined for the LG pairs given its correlation with the
total mass.

We conclude that the typical LG environment and its characteristics is
a robust results depending mostly on the LG total mass.

The alignments with the cosmic web have also a mass
dependency. Although they seem to be stronger once the kinematic constraints
are imposed on the GS pairs, the results are consistent with the
a simple mass cut on the GS sample. There is a clear anti-alignment between
the third eigenvector and the orbital angular momentum vector, meaning
that this vector is perpendicular to filaments and the sheets,  We
also found that the vector joining the LG halos are aligned with the
third eigenvector. This means that he pair is aligned with the filaments and
lies on the sheets and its motion is done along these directions. 

These alignment features are in agreement with the scenario that pairs 
created in-situ or falling into a filament/wall align their orbits with 
the large scale structure in a relaxation process where pair members
tend to moves along the slowest collapsing directions.

How can we relate these alignments to the observed LG? 
To evaluate this point we use observational information for the
satellite distribution around the LG galaxies.  The MW satellites are
located at high galactic latitudes forming a planar structure that
forms an angle of $42^{\circ}-52^{\circ}$ ($\mu=0.6-0.7$) with the
vector joining the MW and M31; while the M31 satellites are on a plane
that lies on the same vector \citep{Pawlowski2013,Shaya2013}. 

Most $\Lambda$CDM studies find that sub-structure infall direction is
done along $\hat{e}_{3}$ (the direction of filaments)
\citep{2005ApJ...629..219Z,2008MNRAS.390.1133B} or almost over the plane
defined by the $\hat{e}_2$ and $\hat{e}_3$ \citep{2014MNRAS.443.1274L}.  

Taking this three points together (the alignment of MW-M31 along
$\hat{e}_3$, the observed satellite alignments and the preferential
infall along $\hat{e}_3$) we have that the M31 plane of satellites is
completely consistent with the average alignment picture we have
described. However, the plane defined by the MW satellites should
also be close to parallel with respect to the vector joining MW and
M31 and the plane of M31 satellites. Instead, this plane has
$\mu=0.6-0.7$ with respect to these directions. Therefore, we suggest
that the spatial location of MW satellites raises a potential
contradiction with the average expectations from $\Lambda$CDM, a point
that has been mentioned before \citep{Pawlowski2012,Pawlowski2014}.  

This apparent contradiction can be solved if one considers that the
subhalo infall properties depend on the environment at {\it the
  merger time}, while the alignments for yet unvirialized pairs such
as the dominant galaxis in the LG depend on the current state of the
cosmic web. This also explain the absence of a strong alignment with
halo spin, because it reflects past configurations of the web, not the
current one. From this perspective, a joint consideration of the
alignments between the dominant halos in the LG and their satellites,
which also feature strong signals
\citep{2005A&A...431..517K,Pawlowski2013,Shaya2013}, should inform us
about the structural evolution of the cosmic web around the LG.   

Our results raise the need to observationally constrain the
alignments of LG pairs with their cosmic web environment. To this end
one could use mass reconstructions from large surveys
\citep{2009MNRAS.394..398W,2011MNRAS.417.1303M,2014MNRAS.445..988N,
  2014ApJ...794...94W} or filament finders
\citep{2010MNRAS.407.1449G,2011MNRAS.414..350S} to select a LG sample
to quantify the alignments with the surrounding filamentary
structure. This would allow not only a direct quantification of how
common are the LG alignments but also provide a new test of
$\Lambda$CDM.

\section*{Acknowledgements}
JEFR was supported by a FAPA grant by Vicerrector\'ia de
Investigaciones at Universidad de los Andes in Bogot\'a, Colombia.
REG was supported by Proyecto Financiamiento Basal PFB-06 'Centro de
Astronomia y Tecnologias Afines' and Proyecto Comit\'e Mixto ESO
3312-013-82. 
The Geryon cluster at the Centro de Astro-Ingenieria UC was
extensively used for the calculations performed in this paper. The
Anillo ACT-86, FONDEQUIP AIC-57, and QUIMAL 130008 provided funding
for several improvements to the Geryon cluster. 
The authors would like to thank Andrey Kravtsov and Nelson Padilla
for their useful comments; Yehuda Hoffman for insightful
conversations on the subject of the Local Group and the Cosmic Web;
and the referee Noam Libeskind  that helped us to clarify and
strenghten the results presented in the paper. 
\bibliographystyle{apj}

\end{document}